\pdfoutput=1
\documentclass[
  aps, prl, twocolumn, superscriptaddress, nofootinbib,floatfix,longbibliography]{revtex4-2}

\usepackage{amsmath,amssymb} 
\DeclareUnicodeCharacter{03C0}{\ensuremath{\pi}}
\DeclareUnicodeCharacter{03C3}{\ensuremath{\sigma}}
\usepackage{braket}
\usepackage{dsfont}

\usepackage{graphicx}
\usepackage{multirow}
\usepackage[normalem]{ulem}
\usepackage{xcolor}
\usepackage[colorlinks=true,linkcolor=blue,urlcolor=blue,citecolor=blue,anchorcolor=blue]{hyperref}
\usepackage[capitalize]{cleveref}
\Crefname{section}{Sec.}{Secs.}
\crefrangelabelformat{equation}{(#3#1#4--#5#2#6)}

\graphicspath{{fig/}}

\newcommand{\tins}{t_{\rm ins}}
\newcommand{\tcut}{t_{\rm ins}^{\rm cut}}
\newcommand{\tslow}{t_s^{\rm low}}
\newcommand{\dEtwo}{\Delta E_{\rm 2pt}}
\newcommand{\dEthree}{\Delta E_{\rm 3pt}}
\newcommand{\dEtwoj}{\Delta E_{\textrm{2pt}, j}}
\newcommand{\dEthreej}{\Delta E_{\textrm{3pt}, j}}

\begin{document}

\title{Nucleon sigma terms in lattice QCD at physical pion mass and in the continuum}

\newcommand{\UCY}{Department of Physics,
University of Cyprus, P.O. Box 20537, 1678 Nicosia, Cyprus}
\newcommand{\CYI}{Computation-based Science and Technology Research Center,
The Cyprus Institute, 20 Kavafi Str., Nicosia 2121, Cyprus}

\author{Constantia Alexandrou} \affiliation{\UCY}\affiliation{\CYI}
\author{Christos Iona}\affiliation{\UCY}\affiliation{\CYI}
\author{Giannis Koutsou} \affiliation{\CYI}
\author{Yan Li} \affiliation{\CYI}
\author{Srijit Paul}\email[Corresponding author: ]{paul.srijit@ucy.ac.cy} \affiliation{\UCY}
\author{Ferenc Pittler} \affiliation{\CYI}

\begin{abstract}
  We determine the light, strange and charm $\sigma$-terms of the nucleon in lattice QCD using four $N_f=2+1+1$ twisted-mass gauge ensembles simulated at physical pion masses and with approximately the same spatial volumes. The four  lattice spacings ranging from  $0.080$~fm to $0.049$~fm, with an action that has $\mathcal{O}(a)$ improvement, enable extrapolating the results to the  continuum limit.  Both valence and sea quark contributions are analysed accounting for excited-state contamination from low-lying multihadron states. We perform global fits to all ensembles,  that yield $\sigma_{\pi N}=53.0(7.4)(1.6)$~MeV, $\sigma_s=58.2(17.0)(2.1)$~MeV and $\sigma_c=112(28)(25)$~MeV, with statistical and systematic uncertainties, respectively.
\end{abstract}
\maketitle

\textit{Introduction.}---The sigma terms of the nucleon, $\sigma_q = m_q \langle N | \bar q q | N \rangle = m_q\, \partial m_N/\partial m_q$, measure how sensitive the nucleon mass is to the quark mass $m_q$ of each quark flavor $q$. They enter the nucleon mass decomposition through the trace of the energy-momentum tensor, together with the gluonic trace anomaly~\cite{Shifman:1978zn,Ji:1994av}. At leading order in the heavy-quark expansion and in $\alpha_s$, the contribution of a heavy quark $Q=c,b,t$ is $\sigma_Q=\tfrac{2}{27}\big(m_N-\sigma_{\pi N}-\sigma_s\big)$, which makes $\sigma_c$ a test of heavy-quark decoupling~\cite{Shifman:1978zn,Hoferichter:2025ubp}. The phenomenological value of $\sigma_{\pi N}$ is extracted from isoscalar pion--nucleon scattering using the Cheng--Dashen low-energy theorem~\cite{Cheng:1970mx,Hoferichter:2015dsa} up to ${\cal O}(m_\pi^4)$. The $\sigma_s$ determines the strangeness content of the nucleon, $y=2m_\ell\sigma_s/(m_s\sigma_{\pi N})$, where $m_\ell$ is the average light-quark mass.

Nucleon sigma terms also connect the scalar quark content of the nucleon to experimental searches for physics beyond the Standard Model. The corresponding mass fractions $f_{T_q}=\sigma_q/m_N$ enter both the Higgs--nucleon coupling and the amplitude for scalar-mediated dark-matter scattering~\cite{Ellis:2008hf,Hoferichter:2017olk}. In particular, they are essential to translate nuclear-recoil limits from LZ, XENONnT, and PandaX-4T into constraints on dark-matter couplings to quarks~\cite{LZ:2024lim,XENON:2023cxc,PandaX:2024qfu}. The same matrix elements enter predictions for Higgs-mediated $\mu\to e$ conversion in the nuclei~\cite{Crivellin:2014cta}.

Lattice QCD determinations of the light-quark sigma term  $\sigma_{\pi N}$ disagree among themselves and with phenomenology. The FLAG average of $N_f=2+1$ results, $42.2(2.4)$~MeV~\cite{FlavourLatticeAveragingGroupFLAG:2024oxs}, lies $3.3\,\sigma$ below the phenomenological value of $56.0(3.5)$~MeV~\cite{Hoferichter:2015dsa,Hoferichter:2023ptl, Alarcon:2011zs}, while the $N_f=2+1+1$ average, $60.9(6.5)$~MeV, agrees with it but rests on only two calculations~\cite{Alexandrou:2014sha, Gupta:2021ahb}. Previous lattice calculations relied largely on a single lattice spacing~\cite{Alexandrou:2019brg} or on joint extrapolations in the lattice spacing and the pion mass~\cite{Yang:2015uis,Gupta:2021ahb,Borsanyi:2020bpd,Agadjanov:2023efe}. An earlier analysis based on ensembles simulated by the Extended Twisted Mass Collaboration (ETMC) at three lattice spacings at physical quark masses agreed with the smaller value~\cite{Alexandrou:2024ozj}. A well-known difficulty in these calculations is the contamination of the nucleon matrix elements by multiparticle excited states involving the nucleon and pion states, which, if not treated properly, can bias the extracted $\sigma_{\pi N}$~\cite{Gupta:2021ahb, Agadjanov:2023efe, Barca:2025det}.

In this work, we  determine $\sigma_{\pi N}$, $\sigma_s$ and $\sigma_c$ using an updated analysis at four lattice spacings with physical quark masses. We take multiparticle states into account when extracting the nucleon matrix elements, and we perform a global fit across the four ensembles in addition to fitting the data from individual ensembles. This enables a more stringent continuum extrapolation, while testing the impact of multiparticle contributions without imposing constraints.

The sigma terms can be determined directly from the forward matrix elements of the scalar quark densities,
\begin{align}
  \sigma_q&\equiv m_q\braket{N(\vec p,s)|\bar qq|N(\vec p,s)}=m_q(\partial m_N/\partial m_q),\label{eq:sigma-def}\\
  \sigma_{\pi N}&=m_\ell\braket{N(\vec p,s)|\bar uu+\bar dd|N(\vec p,s)},\nonumber
\end{align}
with $m_\ell=(m_u+m_d)/2$, $q=s,c$, and the states normalized as $\braket{N(\vec p\,',s')|N(\vec p,s)}=(2\pi)^3\delta^3(\vec p\,'-\vec p)\,\delta_{ss'}$. They are renormalization-group invariant and Lorentz scalar observables. 

The main uncertainty when using exclusively gauge ensembles at physical pion mass comes from the analysis of the contributions by multiparticle states, which carry the quantum numbers of the nucleon. The scalar current excites certain nucleon and pion states strongly while they are barely visible in the determination of the nucleon mass. Thus, analyses that constrain the  excited-state energies from the nucleon mass determination  underestimate $\sigma_{\pi N}$~\cite{Gupta:2021ahb,Gupta:2023cvo,Alexandrou:2024tin,Barca:2025det,Barca:2026juc}.  We allow the effective excited-state gap in the three-point function to differ from that of the nucleon two-point function, and determine it across the four ensembles without priors.

\begin{table}[!htp]
  \caption{Parameters of the ensembles analyzed. The lattice spacings $a$ are from
  Ref.~\cite{ExtendedTwistedMass:2026hvp} (see also Refs.~\cite{ExtendedTwistedMass:2022jpw,ExtendedTwistedMass:2024nyi}). Their error is one on the last digit given. In the last column we give the number of configurations $N_{\rm cfg}$ analyzed per ensemble.
   }
  \label{tab:ensemble}
  \footnotesize
  \begin{ruledtabular}
  \begin{tabular}{lccccc}
  Ensemble   & $(L/a)^3\times T/a$ & $a$ [fm] & $L$ [fm] & $m_N$ & $N_{\rm cfg}$
  \\ \hline
  \texttt{cB211.072.64} (B64) & $64^3\times128$  & 0.0797 & 5.10 & 940(5) &  749 \\
  \texttt{cC211.060.80} (C80) & $80^3\times160$  & 0.0680 & 5.44 & 945(3) &  400  \\
  \texttt{cD211.054.96} (D96) & $96^3\times192$  & 0.0569  & 5.46 & 945(4) &  493 \\
  \texttt{cE211.044.112} (E112) & $112^3\times224$ & 0.0488 & 5.47 & 943(4) &  467 
  \end{tabular}
  \end{ruledtabular}
\end{table}

\textit{Lattice setup.}---We employ four ETMC gauge ensembles with a mass degenerate up and down quarks, a strange and a charm quark  ($N_f=2+1+1$) with their mass fixed to approximately their physical values~\cite{Alexandrou:2018egz,ExtendedTwistedMass:2022jpw,ExtendedTwistedMass:2024nyi}. This lattice discretization scheme at maximal twist gives automatic ${\cal O}(a)$ improvement of physical observables~\cite{Frezzotti:2000nk,Frezzotti:2003ni}, so the leading cutoff effects are ${\cal O}(a^2)$. The  spatial extent  of the ensembles is $L\simeq 5.1$--$5.5$~fm, giving $m_\pi L\simeq3.6$--$3.9$. The parameters of the four ensembles are listed in \cref{tab:ensemble}. Details of the fits are given in a forthcoming paper on the nucleon scalar charges~\cite{Alexandrou:2026scalar}. We use Osterwalder-Seiler~\cite{Osterwalder:1977pc} strange and charm quarks with their bare masses tuned from the meson sector~\cite{ExtendedTwistedMass:2024nyi,ExtendedTwistedMass:2026hvp} at the isospin-symmetric point defined by $m_\pi=135.0$~MeV, $m_K=494.6$~MeV, $m_{D_s}=1967$~MeV and $F_\pi=130.5$~MeV. 

\textit{Correlation functions.}---To extract the nucleon mass, we compute two-point functions $C_2(t_s)$  with the interpolator ${\cal J}_N(t)=\sum_{\vec x}\epsilon_{abc}\big[u^{aT}{\cal C}\gamma_5 d^b\big]u^c(\vec x,t)$ at zero momentum, with Gaussian-smeared quark fields~\cite{Gusken:1989ad,Alexandrou:1992ti} and APE-smeared links~\cite{APE:1987ehd}, tuned to a nucleon smearing radius of about $0.5$~fm~\cite{Alexandrou:2024ozj}. The connected three-point functions $C_3(\tins,t_s)$ are computed with the fixed-sink method, requiring a separate inversion for each value of the time separation $t_s$ between the time of creating and annihilating the nucleon states but for all values of the scalar current insertion times $\tins$. We use a number of source positions per configuration that grows with $t_s$, from one at the smallest to $\mathcal{O}(100)$ at the largest, so that the statistical error is roughly uniform across all values of $t_s$. The disconnected quark loops are computed using the one-end trick, hierarchical probing, color--spin dilution~\cite{Alexandrou:2024ozj, Alexandrou:2026qxc}.

 We form the ratio
\begin{equation}
  R(\tins,t_s)=\frac{C_3^{\rm conn}(\tins,t_s)}{C_2(t_s)}+\frac{C_3^{\rm disc}(\tins,t_s)}{C_2^{\rm high stat.}(t_s)}
 -\braket{S_q}.
  \label{eq:ratio}
\end{equation}
The connected term exists only for $q=u,d$, so $\sigma_s$ and $\sigma_c$ are purely disconnected. Each term is divided by the two-point function computed with the same statistics. The vacuum expectation value of the scalar loop $S_q$ is subtracted. At large time separations $R\to\braket{N|\bar qq|N}$, and $\sigma_q=\mu_q\,R$, with $\mu_q$ the bare quark mass~\cite{ExtendedTwistedMass:2024nyi}. For twisted-mass fermions at maximal twist the product of the bare quark mass and the bare scalar matrix element is renormalization group invariant, so no renormalization is needed~\cite{Alexandrou:2019brg,Alexandrou:2024ozj}. For the light quarks, the scalar current maps onto a flavor nonsinglet current in the twisted basis, so the singlet--nonsinglet mixing that affects $\sigma_{\pi N}$ with Wilson-clover fermions~\cite{Gupta:2021ahb} does not arise.

\textit{Extraction of the nucleon matrix elements.}---With the nucleon at rest and the
positive-parity projector, the correlators have the spectral
decompositions
\begin{align}
  C_2(t_s)&=\sum_i|Z_i|^2\,e^{-E_it_s},\label{eq:spectral}\\
  C_{3}(\tins, t_s)&=\sum_{i,j}Z_iZ_j^*\braket{i|S_q|j}\,
  e^{-E_i(t_s-\tins)-E_j\tins},\nonumber
\end{align}
with $Z_i$ the overlap of the interpolator with the state $i$ of energy $E_i$ with $i=0$ being the nucleon state and $S_q$ the scalar current of flavour $q$. Keeping the nucleon and one effective excited state, whose gap may differ between the three- and the two-point function, the ratio $R_{q,j}$ of \cref{eq:ratio} for flavour $q$ on ensemble $j$, multiplied by the bare quark mass $\mu_{q,j}$, becomes
\begin{align}
  \mu_{q,j}R_{q,j}(\tins,t_s)&=\frac{{\cal C}_{q,j}(\tins,t_s)}{1+c_{1,j}e^{-\dEtwoj(a_j)t_s}},\label{eq:fitglobal}\\
  {\cal C}_{q,j}(\tins,t_s)&=\sigma_{q,j}(a_j)+B_j\,e^{-\dEthreej(a_j)t_s}\nonumber\\
  &\hspace{-4.5em}+A_j\big(e^{-\dEthreej(a_j)\tins}+e^{-\dEthreej(a_j)(t_s-\tins)}\big),\nonumber
\end{align}
where $\sigma_{q,j}(a_j)$ is the sigma term of flavour $q$ at the lattice spacing $a_j$ for ensemble $j$, the overlap $|Z_0|^2$ and the factor $e^{-E_0t_s}$ of the nucleon cancel in the ratio. The energy gaps $\dEtwoj(a_j)$ and $\dEthreej(a_j)$ are the two- and three-point gaps for ensemble $j$. In the denominator, we truncate the spectral decomposition of $C_{2,j}$ to the first excited state, so that $c_{1,j} = |Z_{1,j}/Z_{0,j}|^2$ is the ratio of the overlaps for the first excited state and the nucleon computed on that ensemble. We determine the mass of the nucleon $m_{N,j}(a_j)$ and the energy gap $\dEtwoj(a_j)$ by fitting the effective mass
\begin{equation}
m_{\textrm{eff},\,j}(t)=E_{0,j}(a_j)+\log\frac{(1+c_{1,j}e^{-\dEtwoj(a_j) t})}{(1+c_{1,j}e^{-\dEtwoj(a_j) (t+1)})}
\end{equation}
with $t$ and the energies in lattice units, and then give them as input in \cref{eq:fitglobal} as correlated observables. In the three-point function the overlaps are weighted by $\braket{i|S_q|0}$ (\cref{eq:spectral}), so that another state can dominate, which we expect to be $N\pi\pi$ at rest, as demonstrated by our latest analysis of the B64 ensemble~\cite{Alexandrou:2026kwz} and supported by other studies~\cite{Barca:2025det,Barca:2026juc}. We therefore do not set $\dEthreej(a_j)=\dEtwoj(a_j)$ and fit it together with the other parameters in the numerator of the ratio.
A fit window $(\tslow,\tcut)$ contains the separations $t_s\ge\tslow$ and, at each $t_s$, the insertion times $\tcut\le\tins\le t_s-\tcut$.
From \cref{eq:fitglobal}, it is evident that there are two distinct methods to extract $\sigma_{q}$ in the continuum, one where the extraction is performed per ensemble $j$ and then the \textit{continuum limit} is taken, and one where a \textit{global fit} across all ensembles is performed with a continuum ans\"atze for $\sigma_{q}(a)$ and $\dEthree(a)$.

\textit{Continuum limit.}---We take two approaches:

i) {\it Two-step procedure:} We first fit the ratio of \cref{eq:fitglobal} on each ensemble $j$ separately, at a $\tslow$ chosen on that ensemble where $\sigma_j$ is stable and for two to four values of $\tcut$. The per-ensemble value $\sigma_j$ is the mean over these windows, with the root mean square of the fit errors as statistical error and the standard deviation over the windows as systematic error, as for the global fit below. Since cutoff effects start at ${\cal O}(a^2)$, we take the continuum limit with a term linear in $a^2$ as $\sigma(a)=\sigma+b\,a^2$, by a weighted fit to the $\sigma_j$ of the four independent ensembles. We perform the continuum extrapolation using the statistical uncertainties of the ensemble results $\sigma_j$, propagating them to the continuum value. To estimate the systematic uncertainty from the fit windows, we repeat the continuum extrapolation for every combination of the windows on the four ensembles and take the spread of the continuum values. The statistical and systematic uncertainties are then added in quadrature. The uncertainty of the lattice spacing is propagated and changes the results by less than $0.1$~MeV.

ii) \textit{Global fit.}--Instead of fitting each ensemble separately and then taking the continuum limit, we fit them simultaneously allowing for a linear in $a^2$ dependence in the fit parameters that describe physical quantities, namely the sigma term $\sigma(a)$ and the energy gap $\dEthree(a)$. In \cref{eq:fitglobal} we include this $a^2$ dependence explicitly by setting $\sigma_j=\sigma(a_j)$ and $\dEthreej=\dEthree(a_j)$, where $j$ labels the ensemble, $\sigma(a)=\sigma+b\,a^2$ and $\dEthree(a)=\dEthree+b_E\,a^2$. The energy gap $\dEtwo(a)$ and $c_{1,j}$ are extracted from a correlated two-state fit to the effective masses of all four ensembles for $0.6\le t\le2.2$~fm, with $E_0(a)$ and $E_1(a)$ linear in $a^2$, which gives $m_N=947.1(4.9)$~MeV in the continuum and $\dEtwo=540$--$640$~MeV on the four ensembles; they are used as inputs when fitting the ratio. $\sigma$, $\dEthree$, $b$ and $b_E$ are the only parameters common to all ensembles when fitting the ratio.  The amplitudes $A_j$ and $B_j$ are free on each ensemble, no priors are used, and the covariance is block diagonal because the ensembles are independent. The fit thus extrapolates $\sigma(a)$ to the continuum in one step. %
The global fit uses the same $R$ windows $r=(\tslow,\tcut)$ on every ensemble and gives the continuum value $\sigma_r$ directly in each window. We quote the mean over the windows, $\bar\sigma=\frac1R\sum_r\sigma_r$, with statistical error $\big(\frac1R\sum_r\delta\sigma_r^2\big)^{1/2}$, an upper bound since the windows are correlated, and systematic error $\big(\frac1R\sum_r(\sigma_r-\bar\sigma)^2\big)^{1/2}$.

\begin{figure*}
  \includegraphics[width=\textwidth]{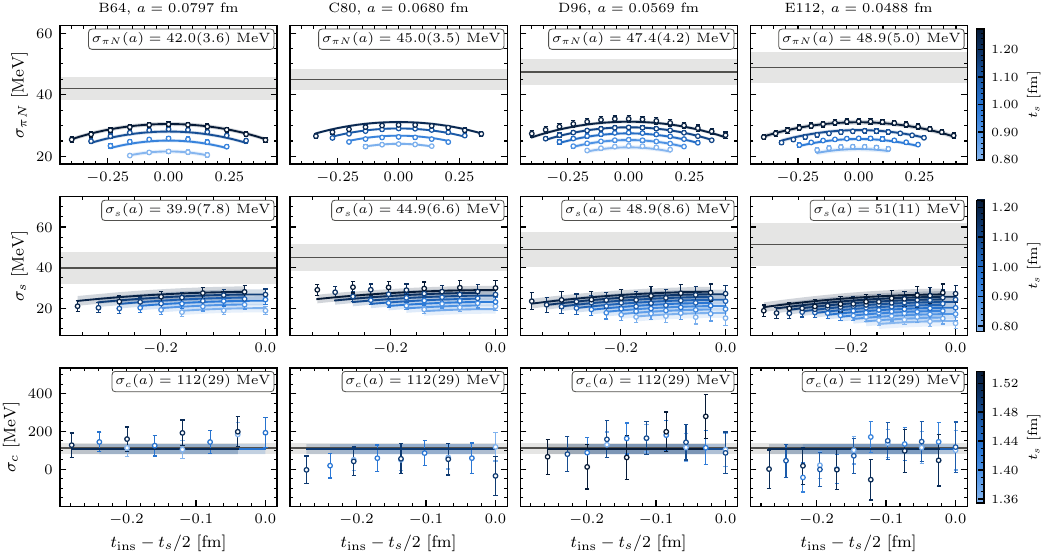}
  \caption{Results on the global fits to the ratios of \cref{eq:ratio}, shown as $\mu_{q,j}R_{q,j}$ times the denominator of \cref{eq:fitglobal}, versus  $\tins-t_s/2$. From left to right we show the fits for the B64, C80, D96 and E112 for  the ratio of  $\sigma_{\pi N}$: (top), for $\sigma_s$ (middle) and $\sigma_c$ (bottom).  The curves and bands are the fits at $\tslow=0.8$~fm and $\tcut=0.24$~fm for $\sigma_{\pi N}$ and $\sigma_s$, and at $1.36\le t_s\le1.54$~fm and $\tcut=0.50$~fm for $\sigma_c$. The gray band and the number in each panel give $\sigma(a_j)=\sigma+b\,a_j^2$ of the global fit at the lattice spacing of that ensemble, averaged over the fit windows; $\sigma_c$ is fitted with a constant, the same on every ensemble.}
  \label{fig:global}
\end{figure*}
\textit{Fits for $\sigma_{\pi N}$.}---We fit the light-quark ratio of \cref{eq:ratio}, with the connected and disconnected contributions combined, by minimizing the fully correlated $\chi^2$, and compute all statistical errors with the $\Gamma$-method~\cite{Wolff:2003sm} in \texttt{pyerrors}~\cite{Joswig:2022qfe}. In both routes the fits start at $\tslow\approx0.8$~fm, below which two states do not describe the data, and use separations up to $t_s\approx1.3$~fm on every ensemble. The fit windows variations in $\tslow$ and $\tcut$ are given in Ref.~\cite{Alexandrou:2026scalar}. The fitted gap, $\dEthree\approx360$--$450$~MeV on the four ensembles, lies close to the lowest noninteracting levels in our volumes, the $N\pi\pi$ threshold at $2m_\pi\approx280$~MeV and $N(\vec p)\pi(-\vec p)$ at $290$--$310$~MeV above $m_N$, as also found in Refs.~\cite{Alexandrou:2026kwz,Gupta:2021ahb,Alexandrou:2024tin}, and well below $\dEtwo\approx540$--$640$~MeV, which is close to the Roper resonance about $500$~MeV above the nucleon~\cite{ParticleDataGroup:2026mpi}. 

\textit{Fits for $\sigma_s$ and $\sigma_c$.}---The purely disconnected strange and charm ratios are symmetrized under $\tins\to t_s-\tins$, so that only the independent half $\tins\le t_s/2$ enters the fits. For $\sigma_s$ we proceed as for $\sigma_{\pi N}$ but set $B_j=0$, since the strange data do not resolve this term. The charm ratio shows no resolvable excited-state curvature, so we fit a constant to the central region of the largest separations; since the four lattice spacings do not resolve an $a^2$ dependence of $\sigma_c$, the linear fit giving $64(89)$~MeV, we take its continuum limit with a constant and add half the difference to the linear fit as a systematic error to the global result. Adding an excited-state term with a free gap, or extending the fits to $t_s=2.0$~fm, shifts $\sigma_c$ by up to one statistical standard deviation.

\begin{figure}
  \includegraphics[width=\columnwidth]{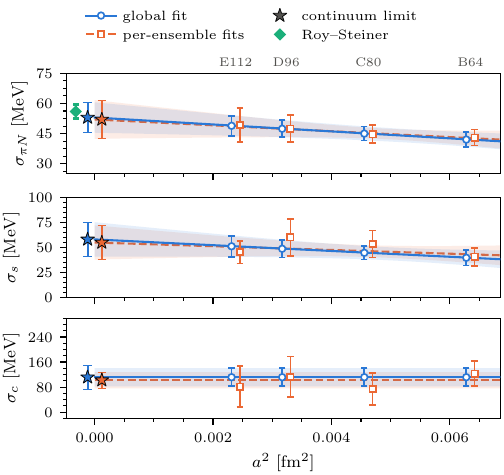}
  \caption{Continuum extrapolation of $\sigma_{\pi N}$, $\sigma_s$ and $\sigma_c$ (top to bottom). Circles: $\sigma(a_j)=\sigma+b\,a_j^2$ of the global fit at the lattice spacing of each ensemble, averaged over the fit windows (constant for $\sigma_c$); squares: the fits of each ensemble separately. Solid (dashed) curves with bands and stars at $a=0$: the continuum extrapolation of the global (per-ensemble) route, linear in $a^2$ except for $\sigma_c$, with statistical and systematic errors added in quadrature. Diamond: the Roy--Steiner value converted to isospin-symmetric QCD~\cite{Hoferichter:2015dsa,Hoferichter:2023ptl}.}
  \label{fig:continuum}
\end{figure}

 {\it Results and discussion.}---
 \Cref{fig:global} shows the global fits together with the ratios of the four ensembles, which they describe with $\chi^2/{\rm dof}\le1.2$ in every window. Even at the largest separations the light and strange ratios lie $30$--$45\%$ below $\sigma(a_j)$, which shows the size of the excited-state contribution removed by the fit, while the charm ratio is flat within its errors. We show the results from the two analysis approaches in
 \cref{fig:continuum}. We quote the global fit results, which extrapolates the ratios to the continuum in one step and takes the correlation between $\sigma(a)$ and the gap into account, and obtain
 \begin{eqnarray}
 \sigma_{\pi N}&=&53.0(7.4)(1.6)~{\rm MeV}=53.0(7.6)~{\rm MeV},\nonumber\\
 \sigma_s&=&58.2(17.0)(2.1)~{\rm MeV}=58.2(17.1)~{\rm MeV},\label{eq:results}\\
 \sigma_c&=&112(28)(25)~{\rm MeV}=112(38)~{\rm MeV},\nonumber
\end{eqnarray}
where the first error is statistical and the second systematic, and the final uncertainty, given last, is the sum of the two in quadrature. For $\sigma_c$ the systematic error includes the $24$~MeV from the unresolved $a^2$ dependence. The $a^2$ slopes of the global fits are compatible with zero; for $\sigma_{\pi N}$, $b=-1.7(1.3)$~GeV\,fm$^{-2}$ amounts to $-21(15)\%$ at the coarsest lattice spacing. Global fits with $\sigma$ and $\dEthree$ constant in $a^2$ give $\sigma_{\pi N}=46.2(4.3)(0.6)$~MeV and $\sigma_s=46.2(6.6)(2.2)$~MeV, and the linear fit without B64 gives $\sigma_{\pi N}=58.9(11.5)(2.5)$~MeV, all within the errors of our results. The finite-volume shift of $\sigma_{\pi N}$ predicted by next-to-next-to-leading-order chiral perturbation theory~\cite{Beane:2004tw,Gupta:2021ahb} with the low-energy constants of Ref.~\cite{Hoferichter:2015hva}, $-1.7$~MeV for B64 and $-1.3$~MeV for the other ensembles, is well within our statistical error and would change the continuum value by about $1$~MeV, so we do not correct for it. The energy gap, $\dEthree=310(72)(21)$~MeV from the light quark sector, agrees within its error with the $N\pi$ and $N\pi\pi$ levels of the four ensembles and is well below $\dEtwo\approx600$~MeV~\cite{Alexandrou:2026scalar}. The per-ensemble route which is the continuum extrapolation of the results of the four ensembles fitted individually, gives $\sigma_{\pi N}=52.0(9.3)(2.5)$~MeV, $\sigma_s=55.1(17.0)(1.8)$~MeV and, with $\sigma_c$ constant, $\sigma_c=102(25)(5)$~MeV, where the first error is statistical and the second the systematic error from the variation over the fit windows.

Since we use the isospin symmetric QCD, defined using  the neutral pion masses, while within the  Roy--Steiner analysis  the isospin limit is defined with the charged-pion mass, we need to convert the latter value to our convention to compare. The value of $\sigma_{\pi N}=59.1(3.5)$~MeV from Ref.~\cite{Hoferichter:2015dsa} is reduced by $3.1(5)$~MeV in isospin symmetric convention, giving $56.0(3.5)$~MeV~\cite{Hoferichter:2023ptl}.  This is the value shown in \cref{fig:continuum} and it is compatible with our value. The covariant chiral perturbation theory analysis of pion--nucleon scattering data of Ref.~\cite{Alarcon:2011zs} gives $\sigma_{\pi N}=59(7)$~MeV, which the same shift reduces to $56(7)$~MeV, also compatible with our value.

\begin{figure}
  \includegraphics[width=\columnwidth]{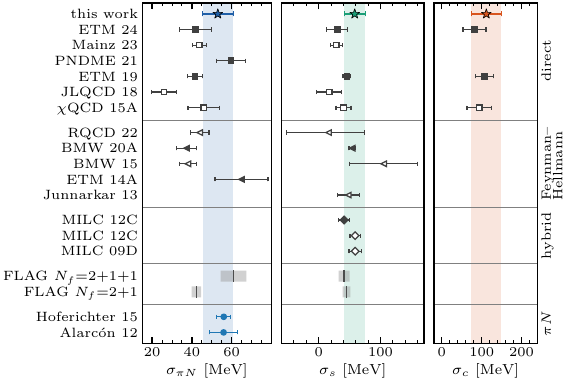}
  \caption{Our continuum results (stars and bands) for $\sigma_{\pi N}$ (left), $\sigma_s$ (middle) and $\sigma_c$ (right) compared with lattice
    determinations from direct matrix elements (squares)~\cite{Alexandrou:2024ozj,Agadjanov:2023efe,Gupta:2021ahb,Alexandrou:2019brg,Yamanaka:2018uud,Yang:2015uis,XQCD:2013odc},
    the Feynman--Hellmann method (triangles)~\cite{RQCD:2022xux,Borsanyi:2020bpd,Durr:2015dna,Alexandrou:2014sha,Junnarkar:2013ac} and the hybrid method (diamonds)~\cite{Freeman:2012ry,Toussaint:2009pz}. We include the  FLAG averages~\cite{FlavourLatticeAveragingGroupFLAG:2024oxs} (gray boxes)
    and the phenomenological determinations from pion--nucleon
    scattering~\cite{Hoferichter:2015dsa,Alarcon:2011zs}, converted to
    isospin-symmetric QCD~\cite{Hoferichter:2023ptl}. Filled (open) symbols use 
    $N_f=2+1+1$ ($N_f=2+1$) ensembles.  Statistical and systematic errors are added in quadrature.}
  \label{fig:comparison}
\end{figure}

In \cref{fig:comparison}, we compare our values with other recent lattice QCD determinations. We distinguish three classes of determinations: those that compute the nucleon matrix elements of the scalar operator, as in this work, those that use the Feynman--Hellmann approach, and those that use the hybrid method of MILC for $\sigma_s$~\cite{Toussaint:2009pz,Freeman:2012ry}. Our values for $\sigma_{\pi N}$ and $\sigma_s$ are in agreement with most lattice determinations except the one from JLQCD which obtained a smaller value for $\sigma_{\pi N}$. Unlike our earlier analysis at three lattice spacings~\cite{Alexandrou:2024ozj}, which gave a smaller $\sigma_{\pi N}$, we keep $\dEthree$ a free parameter and perform a global fit. For $\sigma_c$ there are three other lattice values~\cite{XQCD:2013odc,Alexandrou:2019brg,Alexandrou:2024ozj}, all in agreement with the value extracted from this analysis. The heavy-quark expansion~\cite{Shifman:1978zn} gives at leading order $\sigma_c^{\rm HQE}=\tfrac{2}{27}\big(m_N-\sigma_{\pi N}-\sigma_s\big)\approx61$~MeV, which  agrees with our quoted value. However, the present $34\%$ precision on $\sigma_c$ does  not provide a stringent test of heavy-quark decoupling at the charm mass. 
The results on $\sigma_{\pi N}$ extracted probing the dependence of the nucleon mass on the light quark mass within the Feynman--Hellmann approach tend to give lower values ~\cite{Durr:2015dna,Borsanyi:2020bpd}.

{\it Conclusions.}--- A comprehensive analysis is carried out to extract the light, strange and charm sigma terms of the nucleon   in the continuum limit, using only ensembles at  physical  pion mass and taking into account systematic uncertainties. The analysis includes the accommodation of multiparticle states extracting an energy gap close to the  $N\pi\pi$ threshold, a finding that  is consistent with a variational analysis for B64~\cite{Alexandrou:2026kwz}, where adding nucleon--$\sigma$-meson interpolators and solving the generalized eigenvalue problem~\cite{Blossier:2009kd} substantially reduces the time dependence of the light-quark ratio.
Our values for $\sigma_s$ and $\sigma_c$ provide valuable input for phenomenological searches for beyond the Standard Model physics. For dark-matter direct detection, our $\sigma_{\pi N}$ and $\sigma_s$ give the Higgs--nucleon coupling at leading order in the heavy-quark matching~\cite{Shifman:1978zn,Ellis:2008hf}, $f_N=\tfrac29+\tfrac79(\sigma_{\pi N}+\sigma_s)/m_N=0.314(16)$, in agreement with Ref.~\cite{Hoferichter:2017olk}, where the uncertainty is dominated by $\sigma_s$.

{\it Acknowledgments.}---
We would like to thank all members of the Extended Twisted Mass Collaboration for a very enjoyable cooperation. C.A., C.I., G.K., Y.L., S.P. and F.P. acknowledge partial support from the projects 3D-nucleon (EXCELLENCE/0421/0043), IMAGE-N (EXCELLENCE/0524/0459), MuonHVP (EXCELLENCE/0524/0017), PulseQCD (EXCELLENCE/0524/0269), Baryon8 (EXCELLENCE/0524/0001), RobustSigma (EXCELLENCE/0925/0325), DeNuTra (EXCELLENCE/524/0455), LENA-QCD (POST-DOC/0925 - ROandOther/0180) and partonWF (VISION ERC/0525/0010) funded by the European Regional Development Fund and the Republic of Cyprus through the Cyprus Research and Innovation Foundation. C.A. is supported by European Union's HORIZON MSCA Doctoral Networks programme, under Grant Agreement No. 101072344, project AQTIVATE (Advanced computing, QuanTum algorIthms and data-driVen Approaches for science, Technology and Engineering). The authors gratefully acknowledge the Gauss Centre for Supercomputing e.V. (www.gauss-centre.eu) for funding this project by providing computing time through the John von Neumann Institute for Computing (NIC) on the GCS Supercomputers JUWELS, JUWELS Booster~\cite{JUWELS-BOOSTER} and JUPITER Booster at J\"ulich Supercomputing Centre (JSC). We also acknowledge computing time granted on Piz Daint and Alps at Centro Svizzero di Calcolo Scientifico (CSCS) via the projects s1174 and lp139.

\bibliography{references}

\end{document}